# Large optical nonlinearity enhancement under electronic strong coupling

Kuidong Wang[1], Marcus Seidel[1], Kalaivanan Nagarajan[1], Thibault Chervy[2], Cyriaque Genet[1*], Thomas Ebbesen[1*]

[1]University of Strasbourg, CNRS, ISIS & icFRC, 8 allée Gaspard Monge, 67000 Strasbourg, France. [2]Institute of Quantum Electronics, ETH Zürich, CH-8093 Zürich, Switzerland.

Email: genet@unistra.fr, ebbesen@unistra.fr

## Abstract

**Nonlinear optical responses provide a powerful way to understand the microscopic interactions between laser fields and matter. They are critical for plenty of applications, such as in lasers, integrated photonic circuits, biosensing and medical tools. However, most materials exhibit weak optical nonlinearities or long response times when they interact with intense optical fields. Here, we strongly couple the exciton of organic molecules to an optical mode of a Fabry-Perot (FP) cavity, and achieve an enhancement of the nonlinear complex refractive index by two orders of magnitude compared with that of the uncoupled condition. Moreover, the coupled system shows an ultrafast response of ~120 fs that we extract from optical cross-correlation measurements. The ultrafast and large enhancement of the nonlinar coefficients in this work paves the way for exploring strong coupling effects on various third-order nonlinear optical phenomena and for technological applications.**

Third-order optical nonlinear effects are intrinsic characteristics of a material. Because many significant optical nonlinear phenomena, such as four wave mixing, optical modulation, self-focusing and stimulated Raman scattering, are caused by the third-order optical nonlinear susceptibility[1], they have been explored extensively in various materials ranging from metals[2], semiconductors[3,4,5], 2D materials[6,7], topological insulators[8] to organic materials[9,10,11,12]. An ideal nonlinear optical material should possess large refractive index change at low optical power. In addition, a short response time is also crucial for photonics and optoelectronics applications[7,13]. Usually, the third-order optical nonlinear responses of a material can be described by two parameters, the nonlinear refractive index $n_2$ and nonlinear absorption coefficient $\beta$. These nonlinear coefficients are related to the change in refractive index $\Delta n$ and the modification in attenuation coefficient $\Delta \alpha$ of the material by $n_2 = \Delta n/I$ and $\beta = \Delta \alpha/I$, where $I$ is the intensity of the optical beam[1]. Therefore, one of the central aims in nonlinear optics is to search for or design materials with large $n_2$ or $\beta$. However, standard materials usually show weak nonlinear optical responses even under illumination with strong optical fields. Such limitations therefore call for alternative strategies in order to improve the non-linear responses of existing materials.

One such strategy is to exploit the effect on materials' optical responses of light-matter strong coupling between an excitonic transition and a resonant optical mode of a cavity. When the energy exchange between them is faster than the timescales associated with all dissipative and incoherent processes, two new exciton-polaritonic states are generated, separated in energy by the so-called Rabi splitting (Fig. 1a). Theory shows that such polaritonic states are generated even in the dark due to coupling with vacuum fluctuations of the cavity mode. It has been seen in the past years that the mere presence of such polaritonic states in the coupled system lead to new material properties. For instance, strongly coupled organic molecules could enhance the conductivity[14,15], the rate of energy transfer[16] of molecules, and furthermore, modify the work function[17] and chemical reactions of molecules [18, 19]. Besides, recent studies also showed that second harmonic generation and third harmonic generation could be enhanced in the presence of polaritonic states[20, 21, 22, 23]. However, the measurements in these works did not characterize the intrinsic nonlinear optical parameters such as $n_2$ and $\beta$, which are necessary to evaluate the true potential of strong coupling for all nonlinear optical processes.

In this article, we applied z-scan technique[24] to characterize the nonlinear refractive index and nonlinear absorption coefficient of J-aggregate cyanine molecules that are placed either inside a Fabry-Perot cavity in electronic strong coupling (ESC) condition or outside of it (decoupled situation). As we show below, the formation of the hybrid light-matter states under ESC conditions gives rise to an enhancement of both $n_2$ and $\beta$ values larger than two orders of magnitude. Simulations and modeling show that the large improvement of the nonlinear optical coefficients results not only from the increase of the intracavity electric field at the polaritonic wavelengths, but also, and most remarkably, from an enhancement of the polaritonic dispersion third-order susceptibility itself. In addition, a pulse-width limited ultrafast response (~120 fs) of the coupled system is observed by means of an optical pump-probe measurements. This result demonstrates how ESC can also meet the essential requirements for ultrafast optical modulation and data processing.

**Linear optical measurements**

Our strongly coupled system (ESC cavity) was realized by placing J-aggregates of cyanine molecules dispersed in a polyvinyl alcohol (PVA) polymer inside a planar silver Fabry-Perot (FP). The structure of the molecular monomer is shown in Fig. 1b and all details regarding samples' preparation are given in Supplementary Section S1. As a reference, we use a sample where the same organic material is spin coated on one mirror only, with therefore no possibility of strong coupling. From the linear absorption spectrum (1-T-R) of the bare molecular film, where the center wavelength of the exciton appears at 590 nm (Fig. 1d), we extract the linear refractive index and extinction coefficient by transfer matrix method[25], as indicated in Fig. 1c. The molecular film exhibits a sharp peak in extinction coefficient, which corresponds to a distortion in the linear refractive index of PVA polymer according to Kramers-Krönig relation. Inside the cavity, when the absorption of the molecules is resonant with the optical mode of the cavity, the coupled system as expected, yields two exciton-polaritonic states (noted as |P+> and |P->) at wavelengths of 552 nm and 636 nm, with a Rabi splitting energy of 297 meV. The experimental and simulated results can be seen in Fig. 1d and Supplementary Fig. S2.

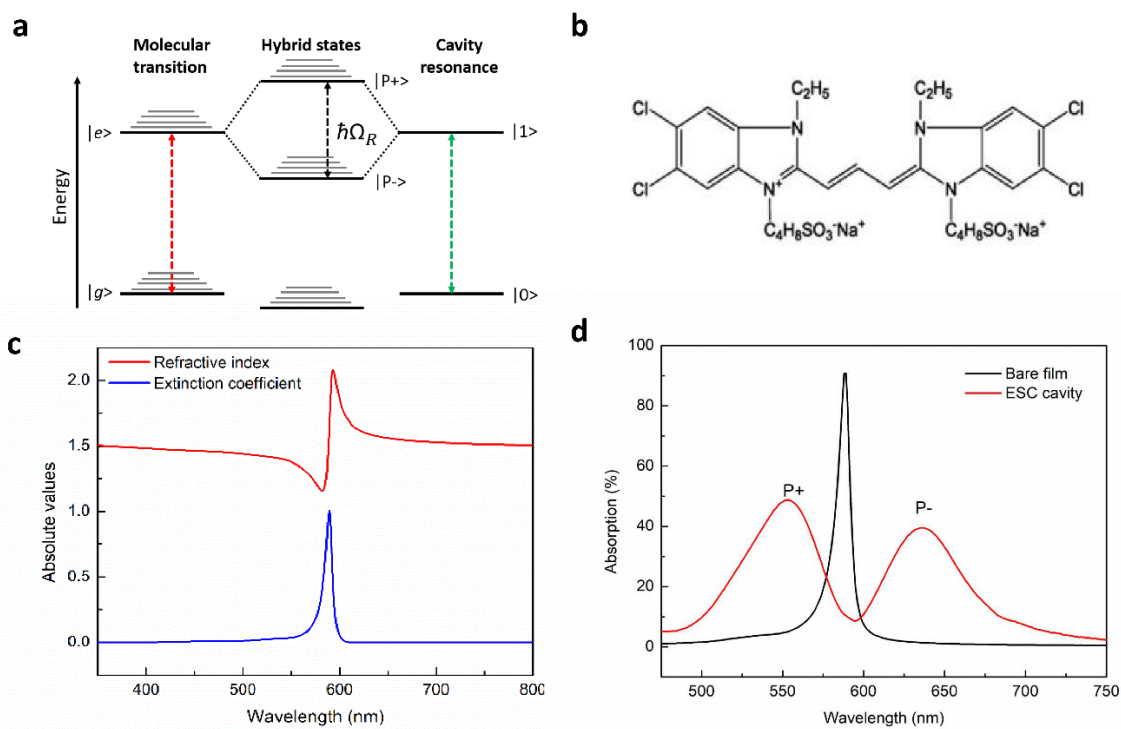

**Fig. 1 | Light-matter strong coupling with organic semiconductors and linear responses of the samples. a**, Schematic energy diagram of the strong coupling between a molecular exciton transition and a cavity resonance. **b**, Chemical structure of the J-aggregating cyanine monomer used in this work that eventually aggregate in the J-form. (See Supplementary Section S1) **c**, Refractive index and extinction coefficient of the bare J-aggregate molecular film calculated by transfer-matrix method. **d**, Linear absorption spectra of the bare J-aggregate molecular film (no cavity - black curve) and of the molecular film under strong coupling (inside the FP cavity - red curve) at normal incidence.

**Nonlinear optical measurements**

We investigated the open- and closed-aperture z-scan measurements in visible range to characterize the nonlinear refractive index and nonlinear absorption coefficient for the ESC cavity and non-ESC sample. The values of $\beta$ can be directly retrieved from the results of open-aperture Z-scan, whereas when analyzing $n_2$ from the closed-aperture z-scan traces, the nonlinear absorption is also taken into account because the energy variations near the center of the transmitted beam stem from both $n_2$ induced extra phase front distortion of the optical beam and the absorption change of the sample near the focus[24]. The transmissive z-scan traces of the molecules inside and outside the cavities for both the open and closed apertures under the irradiance of 625-nm light are displayed in Figs. 2a and 2b, respectively. It should be noted here that the molecules are more easily damaged at the wavelengths near resonances inside the cavity than that under the conditions of outside the cavity. Therefore, in order to avoid this, different levels of incident energy intensities were applied for the different samples. The experimental setup, all parameters used for the laser at the different wavelengths, and the z-scan traces of the open- and closed-aperture measurements are presented in Supplementary Sections S3, S4 and S5, respectively.

This leads us to compare, between the ESC and non-ESC cases, the nonlinear coefficients, $n_2$ and $\beta$, over the full [450 – 750] nm wavelength bandwidth, and in this way to define the enhancement factors of both coefficients as $\eta_{n_2} = (n_2)_{ESC}/(n_2)_{nonESC}$ and $\eta_\beta = (\beta)_{ESC}/(\beta)_{nonESC}$, where $(...)_{ESC}$ and $(...)_{nonESC}$ represent the absolute values of the nonlinear coefficients of the molecules respectively in and out of the strong coupling regime under the same illumination wavelength. As clearly presented with a log scale in Figs. 2c and 2d, compared with the values of $n_2$ and $\beta$ for non-ESC sample, both coefficients are strongly enhanced with a trend that remarkably follows the linear absorption spectrum of the coupled system inside the cavity. In particular, $\eta_{n_2}$ and $\eta_\beta$ reach the maximum values of 122 and 112 near the wavelength of the lower polaritonic state, 625 nm. These two-orders-of-magnitude enhancements result in the values of $-6.49 \times 10^{-10}$ cm²/W and $-3.87 \times 10^{-5}$ cm/W for $n_2$ and $\beta$, respectively, which are more than one order of magnitude larger than those of engineered plasmonic metamaterial[26], and several times higher than those at recently reported for J-aggregate cyanine molecules[27] and that of the nonlinear indium tin oxide in the epsilon-near-zero region[4]. In contrast, smaller than one enhancement factors are measured in between the two polaritonic peaks, i.e. at the middle of Rabi splitting at 590 nm where only dark collective polaritonic states exist that cannot couple with the incident light. A reduction of the nonlinear coefficients also arises in the spectral regions far from the polaritonic peaks where most of the incident light at this wavelength is reflected by the front silver film (Supplementary Fig. S2). Besides, the measured values of $\beta$ for the strongly coupled system have different signs depending on the optical wavelength. This change of sign is due to a laser-induced redshift of the resonances, as detailed in Section S5 of the Supplementary Material.

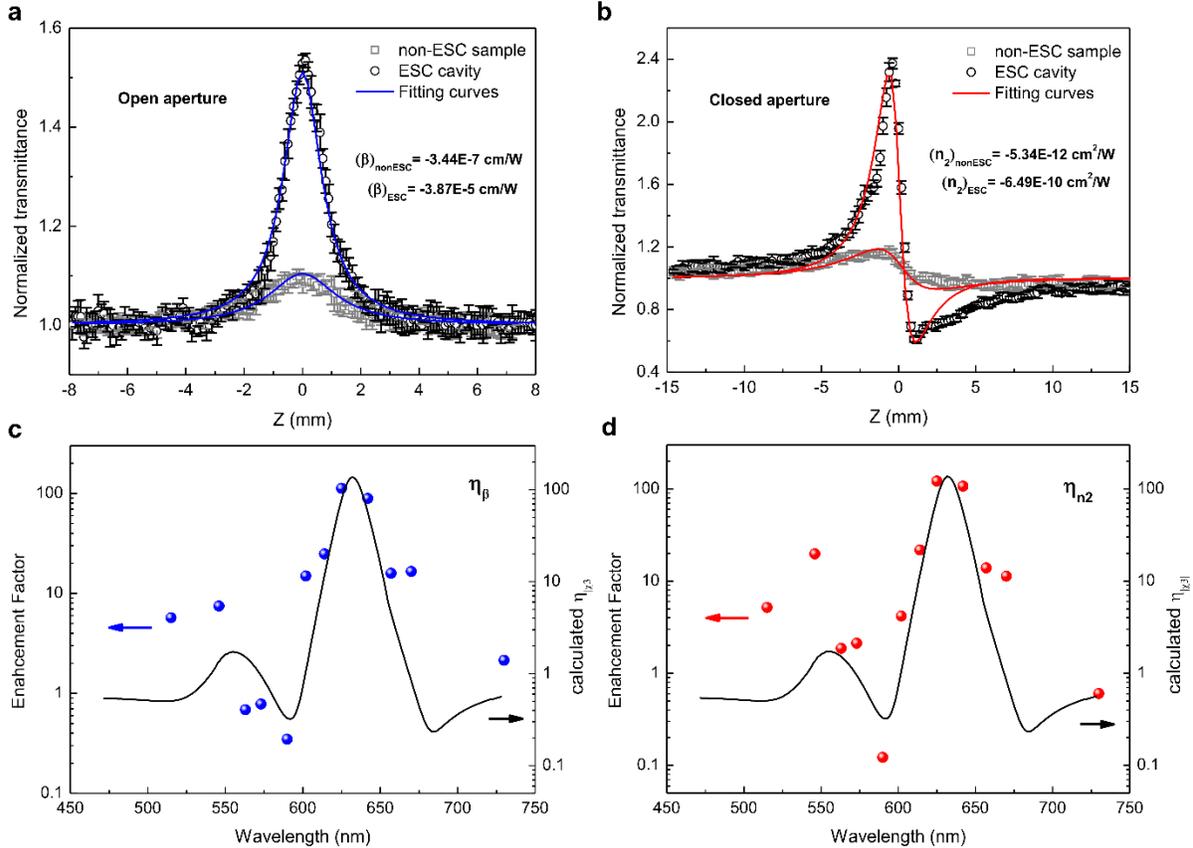

**Fig. 2 | Nonlinear responses of the coupled and uncoupled systems.** Open- (**a**) and closed-aperture (**b**) z-scan traces of the J-aggregate cyanine molecules inside and outside the FP cavity at 625-nm optical illumination. The incident energy for non-ESC sample and ESC cavity are 10 nJ and 0.5 nJ, respectively. The fitting curves in both (**a**) and (**b**) are obtained with the formula that is discussed in Supplementary Section S4. **c**,**d**, are enhancement factors of the retrieved $n_2$ and $\beta$ for various wavelengths. The black curves in (**c**) and (**d**) are the calculated enhancement factor of the absolute values of third-order susceptibility from a simplified nonlinear Lorentz model discussed below.

In order to discriminate and understand the influence of pure cavity resonance on such nonlinear enhancements above, we also carried out z-scan measurements on a thicker molecular film, either placed inside and outside the cavity. The details of these experiments are presented in in Section S6 of the Supplementary Material. For a cavity formed with a thicker film of 193 nm, a cavity resonance is observed at 756 nm, which is far from the wavelength of the J-aggregate excitons. Since in this case, the molecules cannot couple to this cavity resonance, the system is in the weak coupling regime only. At the cavity resonance (756 nm), the enhancement factors of $n_2$ and $\beta$ are only 12 and 5, respectively. This clearly show that nonlinear enhancements merely induced by a cavity effect in this weak coupling regime are much smaller than those measured above in the strong coupling regime. This comparison strongly indicates that the polaritonic states dominate the enhancement in both nonlinear optical coefficients under strong coupling condition.

The temporal response of the molecules under strong coupling was also explored with a degenerate pump-probe measurement at 640 nm. Here, the amplitude of the transmitted probe light is modulated by illuminating a pump beam, with the pulse width of both beams measured at 59 fs. The results displayed in Fig. 3 show an ~120-fs (full width half maximum) peak at time zero, followed by a weak signal with an exponential decay over tens of picoseconds. The zero-delay peak originates in the optical Kerr effect and has a duration proportional to the convolution of the temporal envelope of the probe pulse and the temporal response function of the pump-induced dynamic process in the coupled system. The duration of this Kerr nonlinearity can be even smaller when the pulse width of the pump and probe pulses reduces, indicating it is an upper limit of the intrinsic response time of the strongly coupled system. The temporal response of the slow picosecond component is related to the lifetime of the cyanine molecules[28] and its modulation amplitude is much smaller than that of the Kerr zero-delay peak. The femtosecond time response of the coupled system here is shorter than that of optical nonlinear semiconductors[4, 29, 30], and is thus ideal for ultrafast optical switching applications.

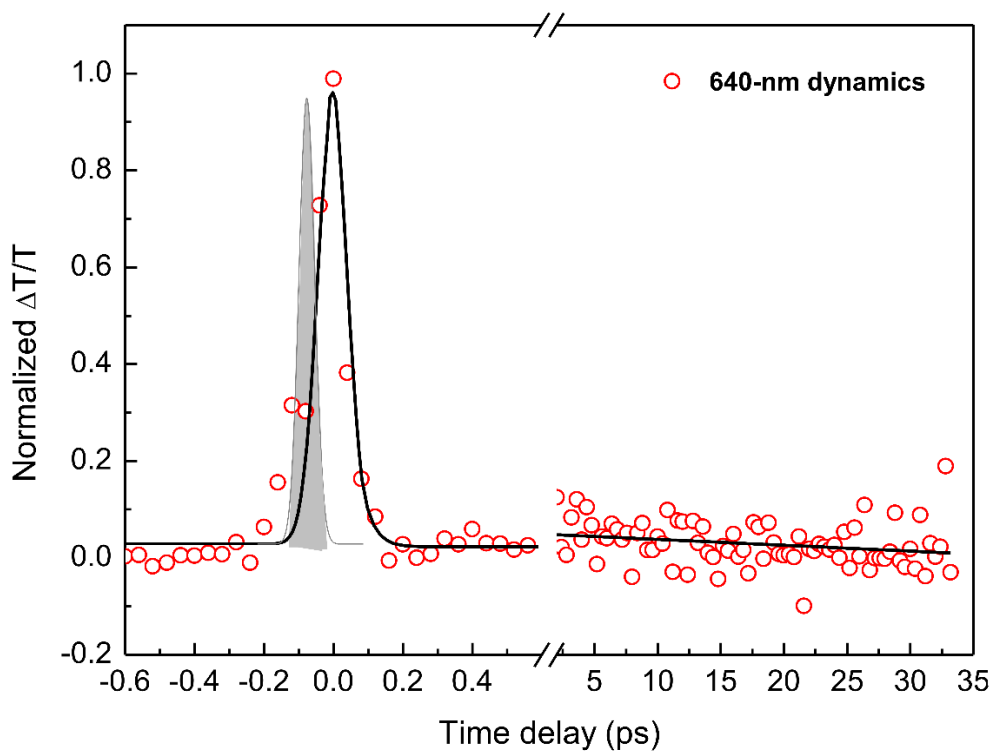

**Fig. 3 | Ultrafast response of the coupled system.** Normalized ultrafast transmittance of the ESC system under a 640-nm optical illumination. The gray area shows the temporal profile of the optical pulse.

As indicated by the results presented above, we attribute the large enhancement values of the nonlinear optical coefficients mainly to the enhancement of the electric field intensity inside the cavity under strong coupling and to the polariton-assisted dispersive third-order susceptibility.

To see this, we first compare the electric field intensity distributions inside the cavity under ESC ($I_{\text{ESC}}$) and outside the cavity -the non-ESC sample ($I_{\text{nonESC}}$)- along the optical z axis. This leads us defining an enhancement factor for the electric field intensity as $\eta_I = I_{\text{ESC}}/I_{\text{nonESC}}$. According to the mean field approximation presented in[31, 32], one can evaluate the enhancement factor of the third-order susceptibility ($\chi^{(3)}$) of the molecules as

$$\eta_{\chi 3} = f \frac{\langle |\mathbf{E}_{\text{ESC}}|^2 \rangle_V \langle \mathbf{E}_{\text{ESC}}^2 \rangle_V}{\langle |\mathbf{E}_{\text{nonESC}}|^2 \rangle_V \langle \mathbf{E}_{\text{nonESC}}^2 \rangle_V} \tag{1}$$

where $\mathbf{E}_{\text{ESC}}$ and $\mathbf{E}_{\text{nonESC}}$ are the electric field distribution within the molecular layer inside and outside the cavity, respectively, and where $\langle ... \rangle_V$ denotes an average of the field intensities taken inside the film over a given volume enclosing a volume fraction $f$ of molecules. Since our molecular film is homogeneous (and composed of only one type of optically active molecules) we fix $f = 1$. Assuming $\langle |\mathbf{E}_\square|^2\rangle_V \langle \mathbf{E}_\square^2\rangle_V \sim I_\square^2$ ($\square$=ESC, non-ESC), the enhancement factor of $\chi^{(3)}$ is directly related to the enhancement factor for the electric field intensity with $\eta_{\chi 3} = \eta_I^2$. This assumption is reasonable because the complex nonlinear refractive index is dominated by its real part in most cases as indicated in Supplementary Table SII. The field intensity enhancement factor can be directly evaluated from the simulations shown in Fig. 4[33]. For our experimental conditions, the comparison between the results of Figs. 4a and 4b gives a maximal intensity enhancement $\eta_I$ distribution along z axis at the lower polariton wavelength. Consider the electric field intensity at 636 nm, the beam size at focal point is much larger than the thickness of the film, thereby the transverse electric field can be regarded as unchanged throughout the pumped unit volume. For the longitudinal electric field intensity, the $I_{\text{ESC}}$ near the middle of the film is ~6, which is much larger than the values of 1 at the two edges, i.e., $\eta_I$ at the central part of the molecular volume dominates the enhancement of nonlinear susceptibility. Hence, when taking into account the $I_{nonESC}$ value at 636 nm (1.5), $\eta_I$ is calculated to be ~4 at lower polariton wavelength and the corresponding $\eta_{\chi 3}$ is ~16. Furthermore, from the optical Kerr effect by a single beam, intensity-dependent complex nonlinear refractive index $\tilde{n}_2$ can be described by[1, 34]

$$\tilde{n}_2 = \frac{3}{4n_0 n_0' \varepsilon_0 c} \chi^{(3)} \qquad (2)$$

where $\tilde{n}_2 = n_2 + in_2'' = n_2 + i\frac{c}{2\omega}\beta$, here $n_0$ and $n_0'$ are the complex and the real part of the linear refractive index, respectively. $n_2''$ is the imaginary part of the nonlinear refractive index, $\varepsilon_0$ and $c$ are the permittivity and the light velocity in vacuum. This indicates that $\tilde{n}_2$ can be enhanced with the same magnitude as $\chi^{(3)}$ when the electric field intensity is boosted under strong coupling condition. Accordingly, as evaluated above, the enhancement factors of $n_2$ and $\beta$ are 16 at the lower polariton wavelength. In addition, the enhancement of the electric field intensity at the lower polariton wavelength is obviously larger than that at the upper polariton wavelength, which is consistent with the enhancement spectra of both $n_2$ and $\beta$ in Figs. 2c and 2d, a trend that confirms that $\eta_I$ contributes to the enhancement of nonlinear coefficients. But remarkably, these 16-times enhancement values remain smaller than the two-orders-of-magnitude enhancement on nonlinear coefficients measured experimentally (Figs. 2c and 2d). This difference points towards other causes for the observed optical nonlinear enhancement under strong coupling.

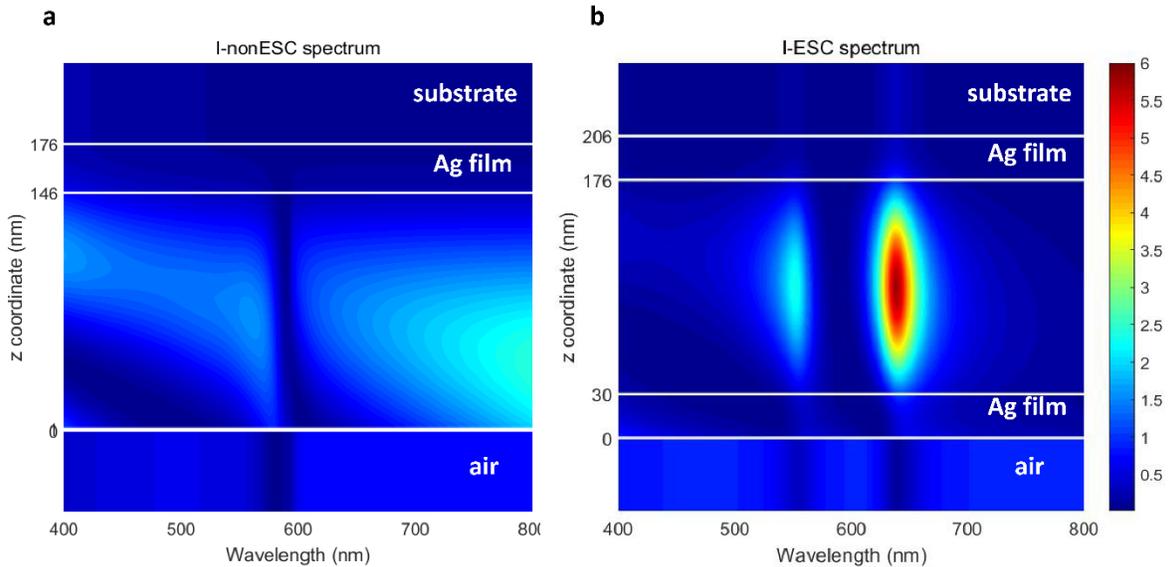

**Fig. 4 | Electric field intensity of the samples. a,b,** Electric field intensity spectra of molecules outside (**a**) and inside (**b**) the FP cavities along z axis.

As already stressed above, the data gathered in Figs. 2c and 2d reveal the signatures of clear resonance enhancements when the frequencies are close to the upper (UP) and lower (LP) polaritonic state energies. Following a simple nonlinear Lorentzian model for the nonlinear third-order susceptibility involving the two polaritonic states and an excitonic state in the coupled system, we can write[35]

$$\chi_{ESC}^{(3)}(\omega) = \sum_{k=UP,LP,ex} \left(\frac{\omega_k^2}{\omega_k^2 - \omega^2 - i\gamma_k\omega}\right)\bar{\chi}_k^{(3)} \tag{3}$$

where $\omega_k$ and $\gamma_k$ correspond to the frequency and linewidth of upper polariton ($k = $ UP), lower polariton ($k = $ LP) and exciton ($k = $ ex), respectively. $\omega$ is the frequency of the pump beam and $\bar{\chi}_k^{(3)}$ is the static third order susceptibilities associated with each excited states involved. Before analyze the dispersive $\chi_{ESC}^{(3)}$ of the coupled system, we should first retrieve the $\bar{\chi}_{ex}^{(3)}$ values of the excitonic state for the uncoupled system, then by using the measured values of $\omega_{UP}$, $\gamma_{UP}$, $\omega_{LP}$ and $\gamma_{LP}$ from the linear absorption spectrum (Fig. 1d) and fitting the best $\chi_k^{(3)}$ values with $\bar{\chi}_{UP}^{(3)} = (0.936 - 0.696i) \times 10^{-17} m^2/V^2$ and $\bar{\chi}_{LP}^{(3)} = -(2.723 + 3.703i) \times 10^{-17} m^2/V^2$, the dispersive nonlinear susceptibility of the coupled system can be well fitted, as illustrated in Supplementary Fig. S6. Furthermore, comparing the calculated dispersive nonlinear susceptibility of the coupled and uncoupled systems, the enhancement of $|\chi^{(3)}|$ directly yields the spectral dispersive features observed in enhancement of the nonlinear coefficients, as presented with black solid curves in Figs. 2(c) and 2(d). This indicates that the polariton-induced dispersive $\chi_{ESC}^{(3)}$ also contributes to the enhancement of the nonlinear optical coefficients. The details of the model and the comparison of the dispersive $\chi^{(3)}$ between coupled and uncoupled systems can be found in Supplementary Section S7. Beside the resonance enhancement effect of the dispersive nature of the polaritonic states, the rather high values for the static third-order susceptibilities could also be related to the large transition moments associated with the delocalized nature of the polaritonic states.

Since both the large electric field intensity inside the cavity at the polaritonic wavelengths and the energy dispersion of $\chi_{ESC}^{(3)}$ at the upper and lower polariton contribute to the significant increase in the nonlinear optical coefficients, the enhancement factor of $n_2$ and $\beta$ can be improved if $(\eta_I)_{\text{UP,LP}}$ and $(\chi^{(3)}_{\text{ESC}})_{\text{UP,LP}}$ are further increased. From the calculations, the $(\eta_I)_{\text{UP,LP}}$ and $(\chi^{(3)}_{\text{ESC}})_{\text{UP,LP}}$ are very sensitive to the quality factor (QF) of the cavity, i.e., strongly increasing with QF. Considering the fact that the QF of the FP cavity is only 12 at 590-nm resonance in our experiments, there is much room to further enhance $(\eta_I)_{\text{UP,LP}}$ and $(\chi^{(3)}_{\text{ESC}})_{\text{UP,LP}}$. For instance, distributed Bragg reflectors[36, 37], which are usually used in polariton condensation studies, and various nanostructures[38] provide possible candidates for high QF cavities. The study of the third-order optical nonlinearity enhancement by exciton-polaritons in this work opens the gates to explore the effect of strong coupling on a series of third-order optical nonlinear phenomena, such as wave mixing, optical modulation and stimulated Raman scattering. In addition, the femtosecond intrinsic response time of the coupled system makes it possible to realize ultrafast optical switching for future applications.

**Conclusion**

In summary, we have performed z-scan measurements on J-aggregate cyanine molecules inside a Fabry-Perot cavity in the electronic strong coupling regime, and measured an enhancement by two orders of magnitude on both the nonlinear refractive index $n_2$ and nonlinear absorption coefficient $\beta$ at the lower polaritonic state compared with measurements done outside the cavity. These large nonlinear enhancements are ascribed to both the increase of the electric field intensity inside the cavity and the polaritonic dispersion of the third-order susceptibility. In addition, we also demonstrated an ultrafast response of ~120 fs of the coupled system using cross-correlation measurements. Such ultrafast, large optical nonlinearities in the strongly coupled system presented here offer an efficient way to realize high speed active photonic and optoelectronic devices.

## Acknowledgements

We acknowledge support of the International Center for Frontier Research in Chemistry (icFRC, Strasbourg), the ANR Equipex Union (ANR-10-EQPX-52-01), the Labex NIE Projects (ANR-11-LABX-0058 NIE), CSC (ANR-10- LABX-0026 CSC), and USIAS (Grant No. ANR- 10-IDEX-0002-02) within the Investissement d'Avenir program ANR-10-IDEX-0002-02, the ERC (project no 788482 MOLUSC) and QuantERA project RouTe. M.S. acknowledges support from the Marie Skłodowska-Curie actions of the European Commission (project 753228, PlaN).

## Author contributions

K.W. and T.W.E. conceived the idea; K.W. and K.N. prepared the samples; K.W. conducted all the optical measurements and the corresponding data analysis with the help of M.S.; M.S., T.C., C.G. and K.W. performed the numerical simulations; All authors contributed to discussing and commenting the manuscript. C.G. and T.W.E. supervised the project.

## Competing interests

The authors declare no competing financial interests.

## Additional Information

Supplementary information is available in the online version of the paper. Reprints and permissions information is available online at …. Correspondence and requests for materials should be addressed to C. G. and T.W.E.

# Supplementary Material for "Large optical nonlinearity enhancement under electronic strong coupling"

Kuidong Wang[1], Marcus Seidel[1], Kalaivanan Nagarajan[1], Thibault Chervy[2], Cyriaque Genet[1*], Thomas Ebbesen[1*]

[1]University of Strasbourg, CNRS, ISIS & icFRC, 8 allée Gaspard Monge, 67000 Strasbourg, France. [2]Institute of Quantum Electronics, ETH Zürich, CH-8093 Zürich, Switzerland.

Email: genet@unistra.fr, ebbesen@unistra.fr

## S1. Sample preparation.

Two types of samples were prepared for our experiments. One sample with molecules enclosed in a cavity (ESC sample) that can be strongly coupled to the cavity mode, and a second sample of molecules without cavity (non-ESC sample), hence uncoupled molecules. The substrates for both samples are 1-mm thick quartz windows. For the ESC sample, a 30-nm silver film was sputtered on the quartz window, before an ~146-nm polymer layer (polyvinyl alcohol, PVA), doped with J-aggregate cyanine molecules TDBC (5,6-Dichloro-2-[[5,6-dichloro-1-ethyl-3-(4-sulfobutyl)-benzimidazol-2-ylidene]-propenyl]-1-ethyl-3-(4-sulphobutyl)-benzimidazolium hydroxide, inner salt, sodium salt, Few Chemicals), was spin coated on top of the silver film. Then, another 30-nm silver film was sputtered on top of the polymer to form a Fabry-Perot cavity, as displayed in Fig. S1. The thickness of the polymer film was carefully selected so as to have the cavity resonance and the molecular absorption peak overlapping. For the non-ESC sample, the thickness of the bottom silver film and TDBC-PVA layer are the same than that of the ESC sample, but without the addition of the top silver film (no cavity formed). The TDBC-PVA solution was made by mixing equal amount of 0.5 wt% TDBC water solution and 5 wt% PVA water solution (molar weight 205000, Sigma-Aldrich).

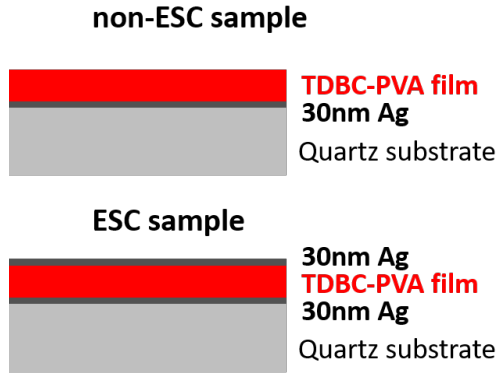

Fig. S1. Schematics of non-ESC and ESC samples.

## S2. Linear response of the strongly coupled system.

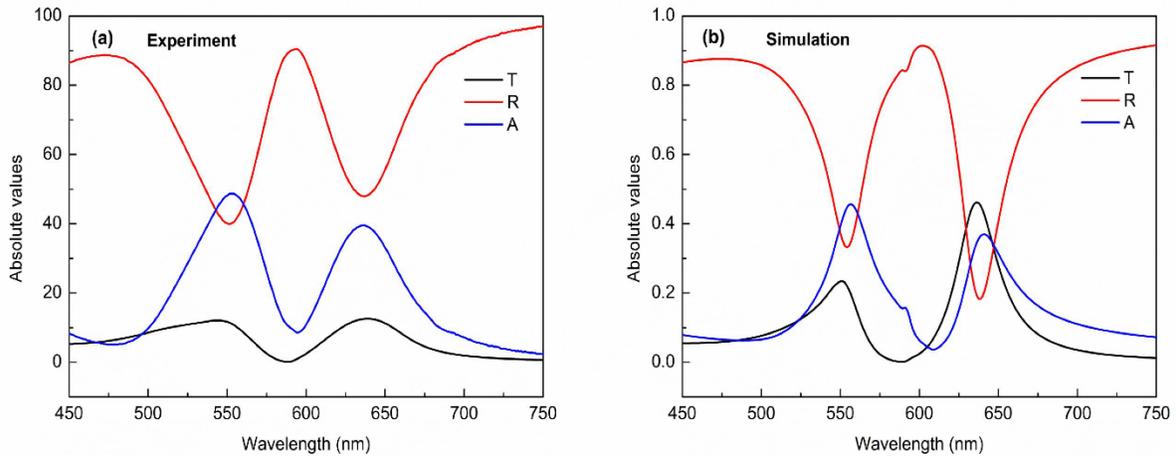

Fig. S2. (a) Measured and (b) calculated linear transmission, reflection and absorption spectra of the strongly coupled system.

The linear transmission, reflection and absorption spectra of the ESC sample were measured at normal incidence using a commercial spectrophotometer, as shown in Fig. S2(a). For all the spectra, two resonances are clearly observed, at wavelengths of 552 and 636 nm, which represent upper and lower polaritonic states. Similar features of the spectra were also obtained by transfer matrix simulation[1]. Here, the simulated amplitudes of the transmission (Fig. S2(b)) at the two resonances are slightly different from that of the measured values, a discrepancy that may stem from the imperfections of the cavity.

## S3. Z-scan setup.

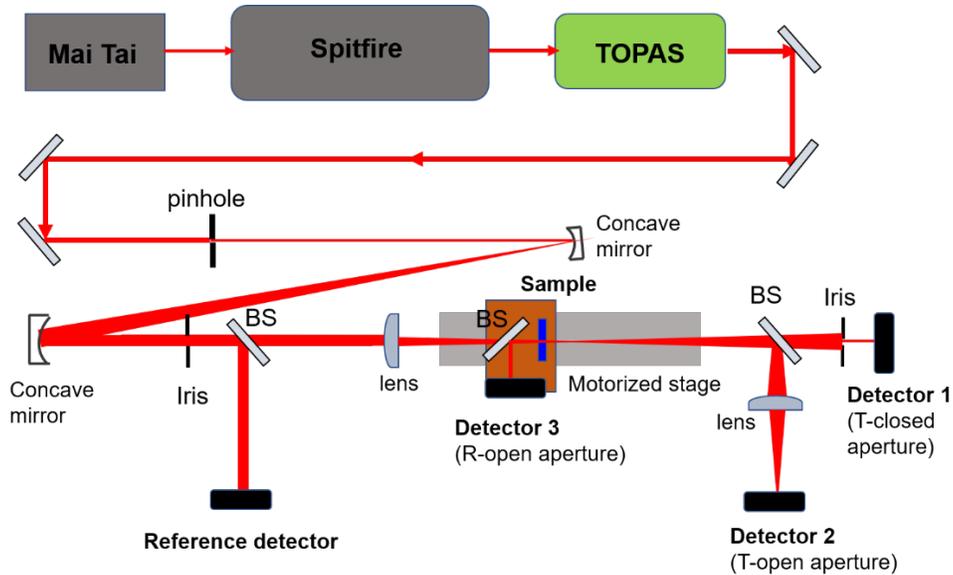

Fig. S3. Schematics of transmissive and reflective z-scan experimental setup.

A schematics of the setup used for z-scan experiments is given in Fig. S3. The tunable visible beams involved in our experiments are generated by an optical parametric amplifier pumped with an 800-nm, 100-fs laser beam generated by a regenerative amplifier (Spitfire, Spectra-Physics). The output light was first passed through a 500-μm pinhole to form a circular symmetric Airy beam. Then, the beam diameter was increased and collimated by a pair of concave mirrors. Diffracted rings far from the optical axis of the beam were trimmed by an iris. This circular beam was finally guided to z-scan setup. Here, the first beam splitter (BS) is used to route one part of the beam to a reference photodetector in order to monitor the energy fluctuations during the experiments. The second and third BSs are used to guide the reflected beam from the sample surface and split the transmitted beam into two portions for the open- and closed-aperture measurements, respectively. The second BS, detector 3 and the sample were mounted on a motorized translation stage. It should be noted that for the reflective z-scan, we just investigated the open-aperture experiments because the size of the reflected beam unavoidable varies when the stage is moving.

## S4. Parameters of the optical beam at various wavelengths.

Table SI. Parameters of the optical beam at different wavelengths

| Wavelength (nm) | Beam size at focal point (μm) | Pulse width (fs) |
|---|---|---|
| 515 | 16.8 | 88.0 |
| 546 | 22.9 | 84.0 |
| 563 | 16.5 | 78.0 |
| 573 | 17.7 | 68.0 |
| 590 | 18.7 | 76.0 |
| 602 | 19.0 | 70.0 |
| 614 | 18.1 | 62.6 |
| 625 | 18.9 | 65.4 |
| 640 | 18.6 | 59.0 |
| 657 | 19.0 | 56.9 |
| 670 | 19.3 | 68.7 |
| 730 | 18.8 | 68.0 |
| 756 | 18.8 | 65.5 |

The beam size at the focal point for the different wavelengths listed in Table SI were measured using a knife-edge method. The corresponding pulse width were obtained with a commercial autocorrelator (<u>Delta Basic Model</u> + UV options, Minioptic Technology, Inc.).

# S5. Z-scan results and optical nonlinear coefficients extraction.

**(a) 545nm**

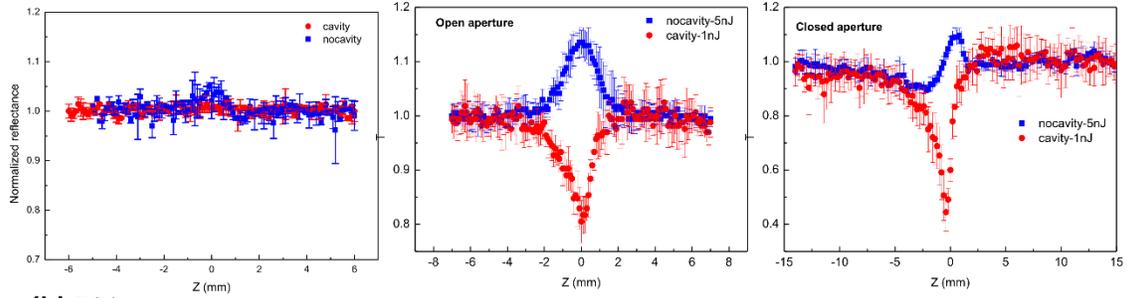

**(b) 590nm**

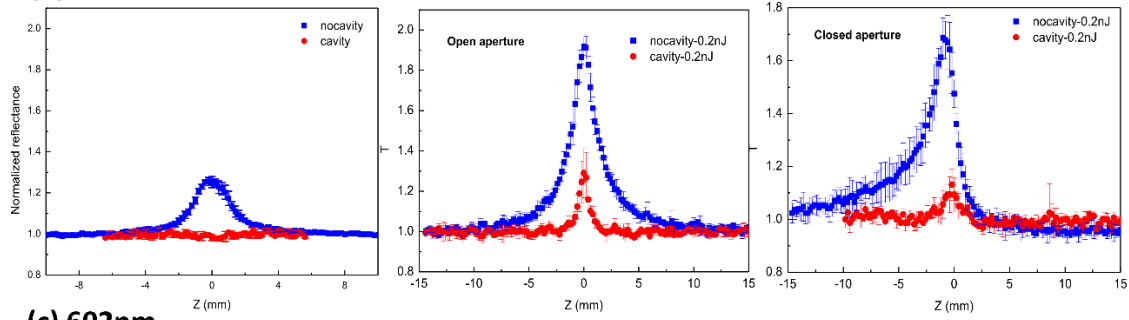

**(c) 602nm**

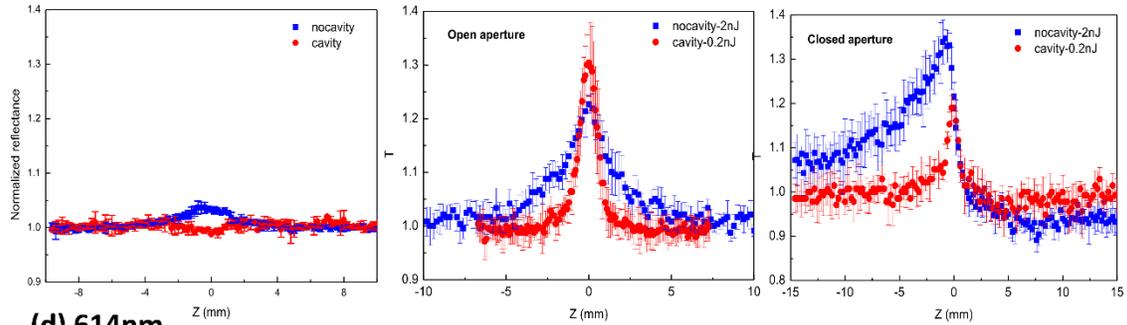

**(d) 614nm**

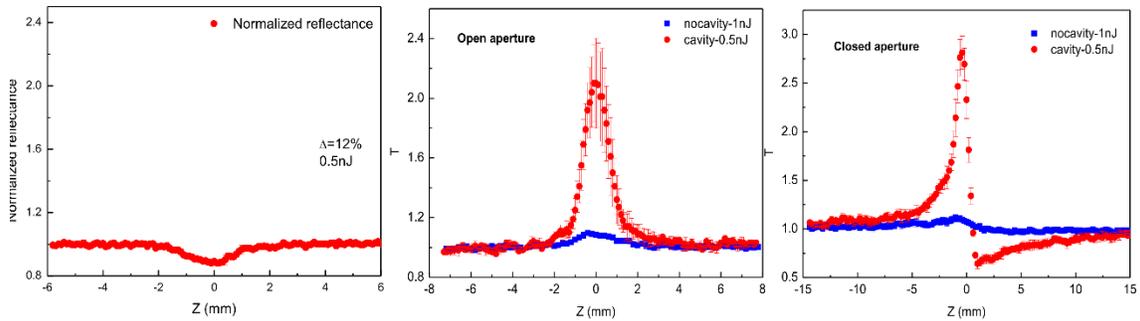

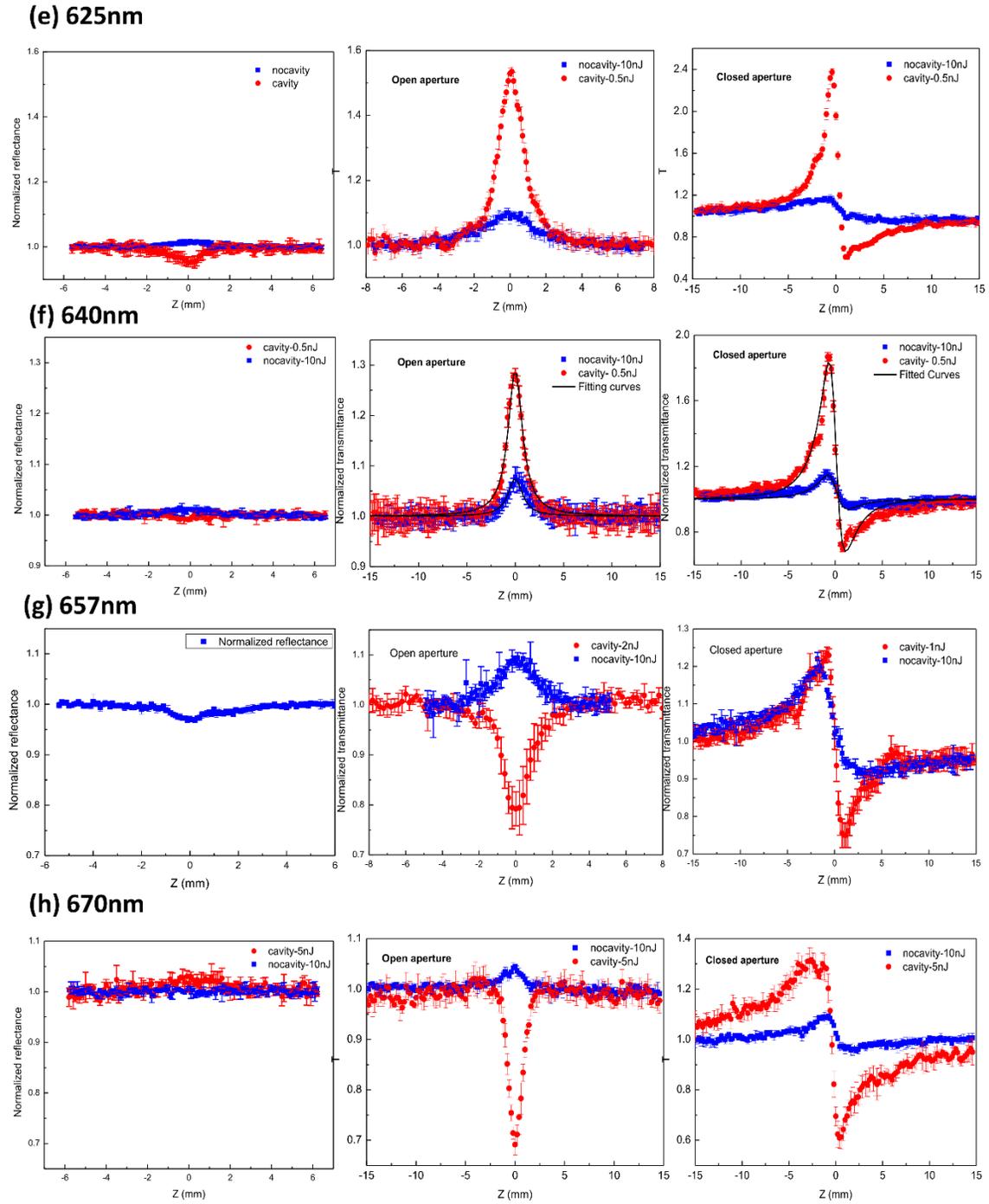

Fig. S4. Open- and closed aperture z-scan traces recorded at several representative wavelengths. Left, middle and right panels of each figure show the results of open-aperture reflective z-scan, open- and closed-aperture transmissive z-scan, respectively. The red and blue traces are obtained from ESC and non-ESC samples, respectively. The error bars of each point are calculated from at least 3 sets of repeated data acquisitions.

The traces of the open- and closed-aperture transmissive z-scan and the open-aperture reflective z-scan at representative wavelengths are displayed in Fig. S4. Obviously, the normalized changes in the reflective z-scan are much smaller than that in the transmitted one in most cases. We therefore extracted the nonlinear absorption coefficient $\beta$ and nonlinear refractive index $n_2$ based on the open- and closed-aperture transmissive z-scan traces. In the case of open aperture, the energy variations within the whole beam profile of the transmitted light were measured. Therefore, modifications in the transmission only relate to the nonlinear absorption of the sample, being strongest at the focus and smallest far from it. A positive peak in the normalized transmission near the focus corresponds to a negative value of $\beta$. In the closed-aperture case, only the energy near the center of the transmitted beam is detected. Thus, variations in transmission are associated to both the nonlinear absorption and nonlinear-refraction induced self-focusing and defocusing.

Assuming a circular profile to the incident beam, the transmission in the open- and closed-aperture configurations can be expressed respectively by[2, 3]

$$T_{OA} = 1 - \frac{\Delta\Psi}{1 + z^2/z_0^2} \tag{S1}$$

$$T_{CA} = \left(1 - \frac{4\Delta\Phi \frac{z}{z_0}}{\left(\frac{z^2}{z_0^2} + 9\right)\left(\frac{z^2}{z_0^2} + 1\right)}\right) \tag{S2}$$

where $\Delta\Psi = \beta I_{peak} L_{eff}/2^{3/2}$, $\Delta\Phi = 2\pi n_2 I_{peak} L_{eff}/\lambda$, and $z$, $z_0$, $I_{peak}$, $L_{eff}$ are the sample position, the Rayleigh range of the beam, the optical intensity at the focus, the effective length, respectively. The effective length and the linear absorption coefficient $\alpha$ are related by $L_{eff} = \frac{1-e^{-\alpha L}}{\alpha}$, where $L$ is the sample thickness. Therefore, the nonlinear coefficients $n_2$ and $\beta$ can be obtained by fitting the experimental z-scan results with the equations above.

As illustrated in Fig. S4, we utilized different optical energies for various optical wavelengths to avoid any damage of the samples under pulsed illumination, meanwhile achieving measurable variations in transmission. For the results of the open aperture, the values of $\beta$ are negative (positive) when the transmission is enhanced (suppressed) near the focus. As shown in the middle panel of each figures in Fig. S4, the molecules exhibit negative $\beta$ at all wavelengths outside the cavity (non-ESC sample). In contrast for the ESC case, the nonlinear absorption coefficients show different signs when changing the optical pump wavelengths. Considering the linear transmission spectrum of the ESC sample, we think that the wavelength-dependent sign of $\beta$ results from the laser induced redshift of the resonances. It should be noted here, that during the extraction of the nonlinear coefficients of the ESC and non-ESC samples, we do not take into account the nonlinear effects of the quartz substrate and silver film, because they yield very small values for $n_2$ and $\beta$[3,4], and the variations in their transmission cannot be resolved with our illumination maximum energy.

## S6. Z-scan measurements of the reference cavity without strong coupling.

In order to understand the pure cavity effect on the enhancement of the optical nonlinear coefficients, we also prepared reference samples with 193-nm thicker TDBC-PVA film inside and outside the cavities. As presented in Fig. S5(a), the wavelength of the cavity resonant mode was then set to 756 nm, which is far from the wavelength of excitons. Therefore, these reference samples cannot induce any strong coupling. After extracting the nonlinear coefficients of the reference samples, we found that the enhancement factors of $n_2$ and $\beta$ are only 12 and 5, respectively, which are much smaller than the enhancements measured at the polaritonic states.

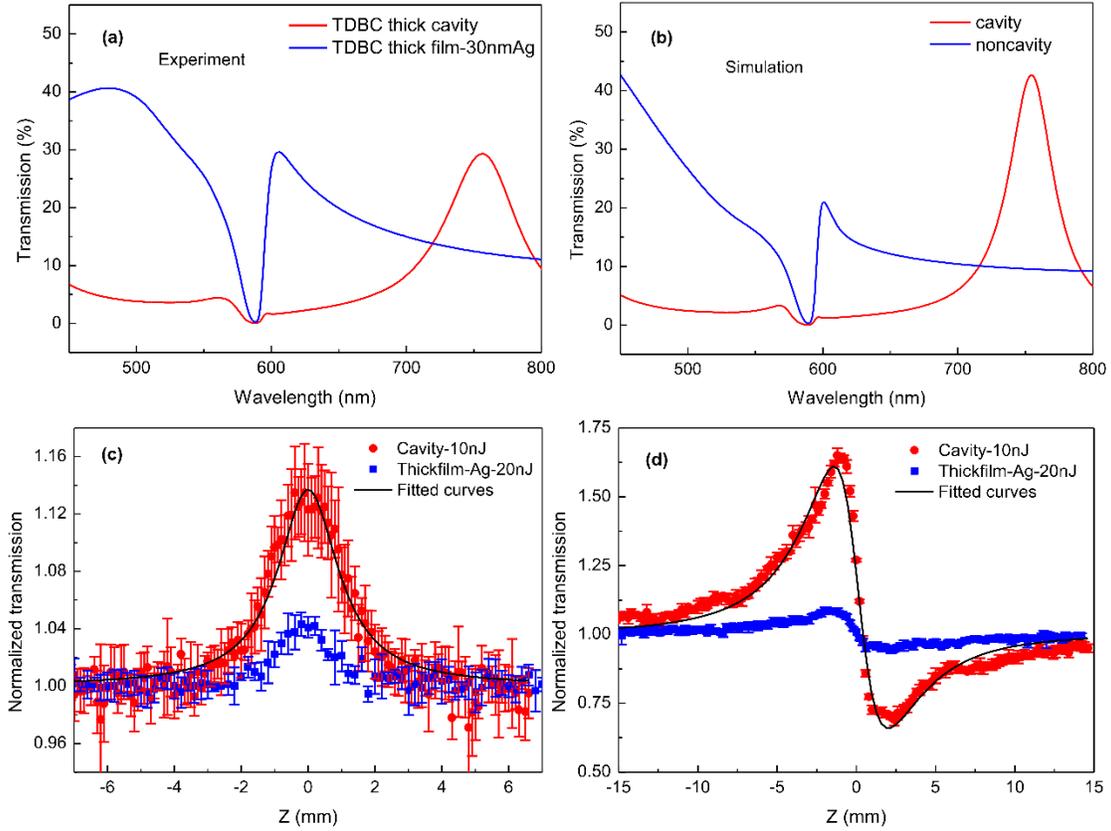

Fig. S5. (a) Experimental and (b) simulated linear transmission spectra of the thick TDBC-PVA film inside (red curve) and outside (blue curve) the cavities. Open- (c) and closed-aperture (d) transmissive z-scan traces on thick TDBC-PVA films inside and outside the cavity, respectively.

## S7. Simplified model on third-order nonlinear susceptibility.

In order to correlate the third-order nonlinear susceptibility and the formation of polaritonic states in strongly coupled systems, we consider a simplified nonlinear Lorentz model. In its simple formulation, the nonlinear Lorenz model consider the time evolution of the polarization density $P$ of the system as damped harmonic oscillator with[5]

$$\frac{d^2 P}{dt^2} + \gamma \frac{dP}{dt} + \omega_0^2 P = \omega_0^2 \varepsilon_0 \{\bar{\chi}^{(1)} E + \bar{\chi}^{(2)} E^2 + \bar{\chi}^{(3)} E^3 + \cdots\} \qquad (S3)$$

where $\gamma$ and $\omega_0$ represent the linewidth and frequency of the transition, respectively. $\bar{\chi}^{(1)}$, $\bar{\chi}^{(2)}$ and $\bar{\chi}^{(3)}$ correspond to linear, second-order and third-order static susceptibilities, respectively. Using the Fourier transform and assuming that each transition is associated with an independent oscillator, we can get the third-order dispersion equation from the equation (S3) as

$$\chi^{(3)}(\omega) = \sum_k \left(\frac{\omega_k^2}{\omega_k^2 - \omega^2 - i\omega\gamma_k}\right) \bar{\chi}_k^{(3)} \qquad (S4)$$

where $\omega_k$, $\gamma_k$ and $\bar{\chi}_k^{(3)}$ are the frequency, linewidth and third-order static susceptibility of the $k$ oscillator, respectively. For the uncoupled system, where only excitons exist, we can replace the subscript $k$ by ex only, representing the parameters of the excitonic oscillator. Similarly, for the coupled system, we use UP and LP to describe the parameters associated with the upper and lower polaritonic oscillators.

Following this model, the dispersive susceptibility of the uncoupled and coupled systems studied in our experiments can be respectively expressed by

$$\chi_{\text{nonESC}}^{(3)}(\omega) = \left(\frac{\omega_{\text{ex}}^2}{\omega_{\text{ex}}^2 - \omega^2 - i\omega\gamma_{\text{ex}}}\right) \bar{\chi}_{\text{ex}}^{(3)} \qquad (S5)$$

$$\chi_{\text{ESC}}^{(3)}(\omega) = \left(\frac{\omega_{\text{UP}}^2}{\omega_{\text{UP}}^2 - \omega^2 - i\omega\gamma_{\text{UP}}}\right) \bar{\chi}_{\text{UP}}^{(3)} + \left(\frac{\omega_{\text{LP}}^2}{\omega_{\text{LP}}^2 - \omega^2 - i\omega\gamma_{\text{LP}}}\right) \bar{\chi}_{\text{LP}}^{(3)} \qquad (S6)$$

$$+ \left(\frac{\omega_{\text{ex}}^2}{\omega_{\text{ex}}^2 - \omega^2 - i\omega\gamma_{\text{ex}}}\right) \bar{\chi}_{\text{ex}}^{(3)}$$

Based on equation (2) in the main text, we can calculate the values of both the complex nonlinear refractive index $\tilde{n}_2$ (Table SII) and complex third-order nonlinear susceptibility (red and blue dots in Fig. S6) from the results of the z-scan measurements. For the uncoupled system, only the excitonic oscillator contributes to the dispersive nonlinear susceptibility. Here, we first decompose equation (S5) into real and imaginary parts, then use the linear values of $\omega_{ex} = 2.1$ eV, $\gamma_{ex} = 0.032$ eV (Fig. 1d in the main text) and fit the corresponding parts of $\chi^{(3)}_{nonESC}$, as presented in Figs. S6a and S6b, in order to obtain the best fitting $\bar{\chi}^{(3)}_{ex}$ value of $(10.01 - 1.943i) \times 10^{-17}$ m²/V². The acquisition of $\bar{\chi}^{(3)}_{ex}$ allows us to further analyze the contribution of the polaritonic states to the dispersive nonlinear susceptibility of the coupled system. For the coupled system, the polaritonic parameters $\omega_{UP}$, $\gamma_{UP}$, $\omega_{LP}$ and $\gamma_{UP}$ can be extracted from the linear absorption spectrum (Fig. 1d) to be 2.25, 0.201, 1.95 and 0.141 eV, respectively. With these values, the real and imaginary parts of the dispersive susceptibilities of the coupled system can be well fitted, as shown in Figs. S6c and S6d, giving the fitting values for $\bar{\chi}^{(3)}_{UP} = (0.936 - 0.696i) \times 10^{-17}$ m²/V² and $\bar{\chi}^{(3)}_{LP} = -(2.723 + 3.703i) \times 10^{-17}$ m²/V². With this fitted nonlinear Lorenz model, we can eventually calculate the polariton-induced dispersive nonlinear susceptibility and the corresponding enhancement factor of the absolute values of $\chi^{(3)}$ ($\eta_{|\chi 3|}$) from the fitting curves in Fig. S6, which can be found in Figs. 2c and 2d in the main text.

Table SII. Extracted $\tilde{n}_2$ from the z-scan measurements (unit: $10^{-16}$ m²/W)

| Wavelength (nm) | non-ESC sample | ESC sample |
| --- | --- | --- |
| 515 | 0.046+0.019i | 0.24+0.11i |
| 546 | 0.093-0.054i | 1.78+0.46i |
| 563 | 1.25+2.23i | 2.46+1.61i |
| 573 | 1.22+3.10i | 3.03+2.67i |
| 590 | 3.93-67.1i | 4.89-20.4i |
| 602 | -0.615-0.255i | -2.52-3.70i |
| 614 | -0.261-0.141i | -5.65-3.50i |
| 625 | -0.049-0.016i | -5.90-1.76i |
| 640 | -0.040-0.009i | -4.23-0.83i |
| 657 | -0.045-0.010i | -0.62+0.16i |
| 670 | -0.039-0.007i | -0.44+0.12i |
| 730 | -0.029-0.006i | -0.018+0.012i |

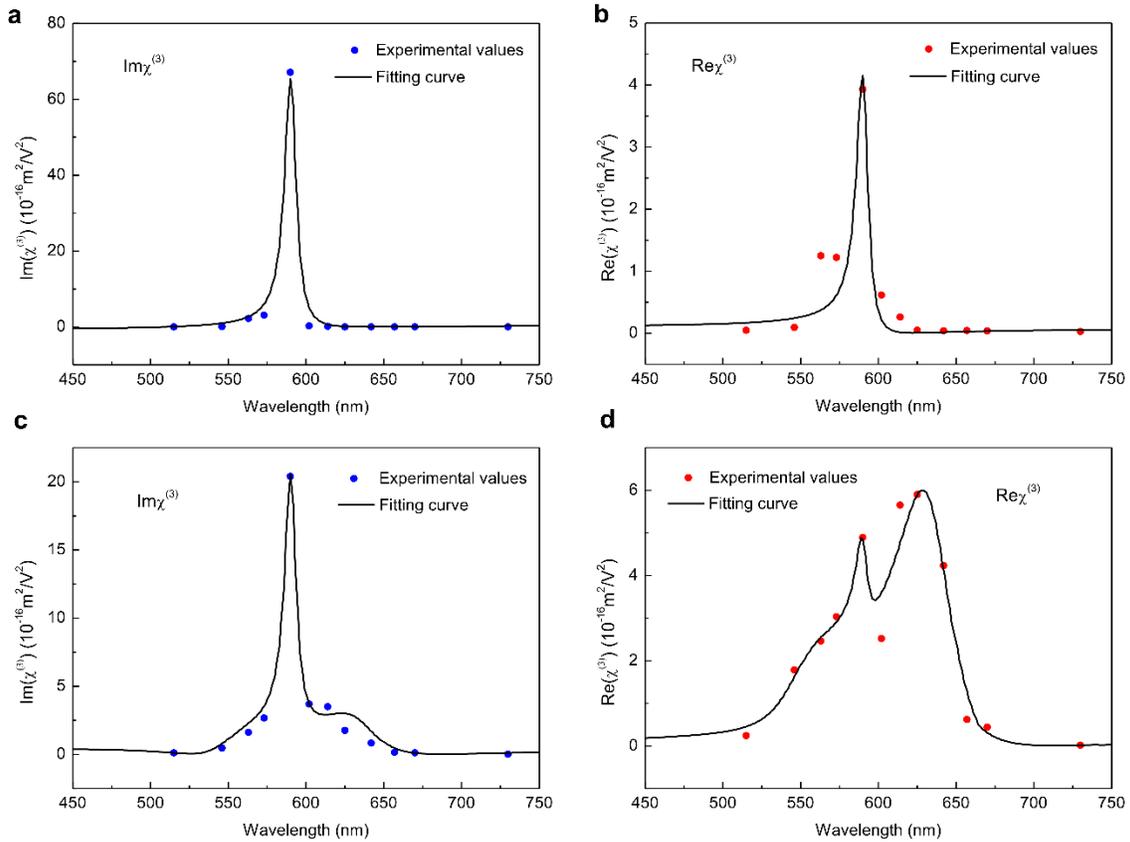

Fig. S6. Experimental and fitted (nonlinear Lorentz model) absolute values of the third-order susceptibility of the uncoupled (a, b) and coupled (c, d) systems. The blue dots in a, c and red dots in b, d represent the imaginary and real parts of the nonlinear susceptibility.